\documentclass[epj]{webofc}
\usepackage[utf8]{inputenc}
\usepackage[varg]{txfonts}   
\usepackage{booktabs}
\usepackage{xcolor}
\definecolor{darkred}{rgb}{0.4,0.0,0.0}
\definecolor{darkgreen}{rgb}{0.0,0.4,0.0}
\definecolor{darkblue}{rgb}{0.0,0.0,0.4}
\usepackage[bookmarks,linktocpage,colorlinks,
    linkcolor = darkred,
    urlcolor  = darkblue,
    citecolor = darkgreen]{hyperref}
\usepackage{subfigure}
\wocname{EPJ Web of Conferences}
\woctitle{Lattice2017}
\begin{document}
%
\selectlanguage{english}
\title{%
SU(3) sextet model with Wilson fermions
}
\author{%
\firstname{Martin} \lastname{Hansen}\inst{1} \and
\firstname{Claudio}  \lastname{Pica}\inst{1}\fnsep\thanks{Speaker, \email{pica@cp3.sdu.dk}}
}
\institute{%
CP$\mathit{\, ^3}$-Origins, University of Southern Denmark, Campusvej 55, DK-5230 Odense M, Denmark
}
\abstract{%
  We present our final results for the SU(3) sextet model with the non-improved Wilson fermion discretization. We find evidence for several phases of the lattice model, including a bulk phase with broken chiral symmetry. We study the transition between the bulk and weak coupling phase which corresponds to a significant change in the qualitative behavior of spectral and scale setting observables. In particular the $t_0$ and $w_0$ observables seem to diverge in the chiral limit in the weak coupling phase. We then focus on the study of spectral observables in the chiral limit in the weak coupling phase at infinite volume. We consider the masses and decay constants for the pseudoscalar and vector mesons, the mass of the axial vector meson and the spin-1/2 baryon as a function of the quark mass, while controlling finite volume effects. We then test our data against both the IR conformal and the chirally broken hypotheses.\\[.1cm]
  {\footnotesize  \it Preprint: CP3-Origins-2017-49 DNRF90  }  
}
\maketitle
\section{Introduction}\label{intro}
The discovery of the Higgs boson, with properties closely resembling the SM elementary Higgs, excludes a large number of BSM models, such as the ``Higgsless models'', and the traditional Technicolor theories, based on QCD-like dynamics. However, a wide class of composite Higgs theories, in which EW symmetry is broken dynamically by a new strong force, are still compatible with the experiments.

The two most interesting realizations of composite Higgs models are Walking Technicolor (WTC) and pNGB Higgs models. 
In such models the Higgs is regarded as the pseudo Nambu-Goldstone boson (pNGB) of an approximate global symmetry, which explains the little hierarchy between the mass of the Higgs and the other resonances of the strong sector. This extra symmetry is a global flavor symmetry in the case of pNGB Higgs models, or an approximate scale invariance symmetry in the case of Walking Technicolor.

Walking Technicolor models are asymptotically free models which can be considered as a small continuous deformation of a conformal field theory in the infrared. Such models share several important features, which makes them good candidates for a composite Higgs model. This includes the possible emergence of a light $0^{++}$ scalar state, associated with the approximate scale invariance, that can play the role of the Higgs boson, with a light mass and couplings similar to the SM Higgs. The strong coupling might evolve slowly with the energy scale (i.e. it \textit{walks}), and if the model also has a large mass anomalous dimension $\gamma\sim1-2$, SM fermion masses could be generated without large Flavor Changing Neutral Currents. Furthermore, models with only two EW gauged fermions are favored as they have a smaller $S$-parameter and do not violate constraints on EW precision tests \cite{Sannino:2004qp}.

Here we study on the lattice the so-called ``sextet model", a WTC model based on an SU(3) gauge theory with a doublet of Dirac fermions in the two-index symmetric (sextet) representation. 

The sextet model has been studied previously on the lattice by several groups \cite{Shamir:2008pb, DeGrand:2008kx, DeGrand:2010na, Kogut:2010cz, Kogut:2011ty, DeGrand:2012yq, Kogut:2015zta, Hasenfratz:2015ssa, Fodor:2015zna, Fodor:2012ty, Fodor:2014pqa, Fodor:2016wal, Fodor:2016pls, Fodor:2015eea}, by using different lattice discretization and many groups focused on the computation of the non-perturbative $\beta$-function in some given scheme.

Recent studies of the spectrum with improved rooted staggered fermions point to the model being chirally broken \cite{Fodor:2012ty,Fodor:2014pqa,Fodor:2016pls,Fodor:2015eea}.

Here we present our final results for the spectrum obtained by using the Wilson discretization, now published in \cite{Hansen:2017ejh} and presented in a partial form at previous Lattice conferences \cite{Hansen:2016sxp,Drach:2015sua}.

\begin{figure}
  \begin{center}
    \includegraphics[height=5cm]{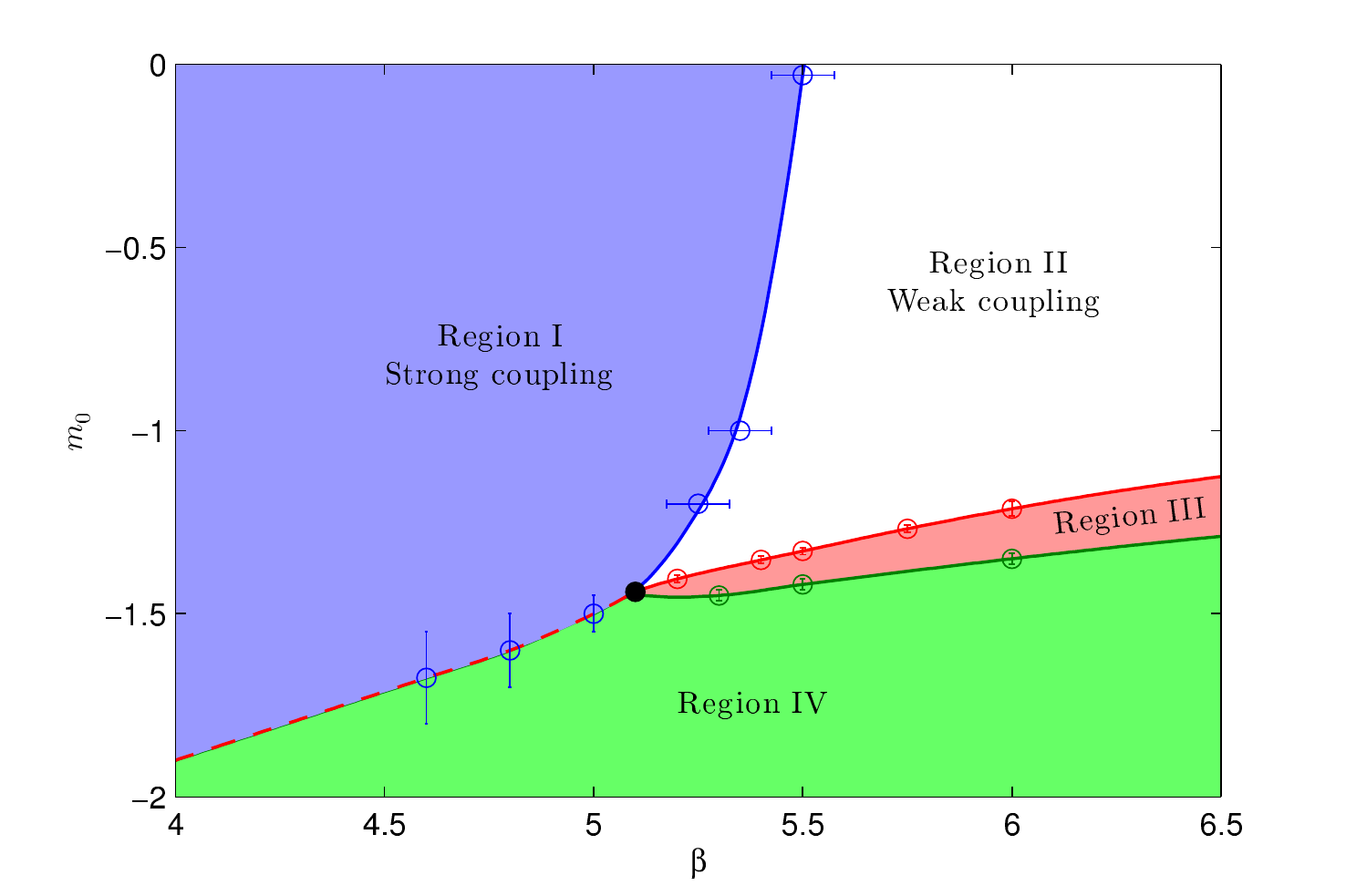}
    \includegraphics[height=5cm]{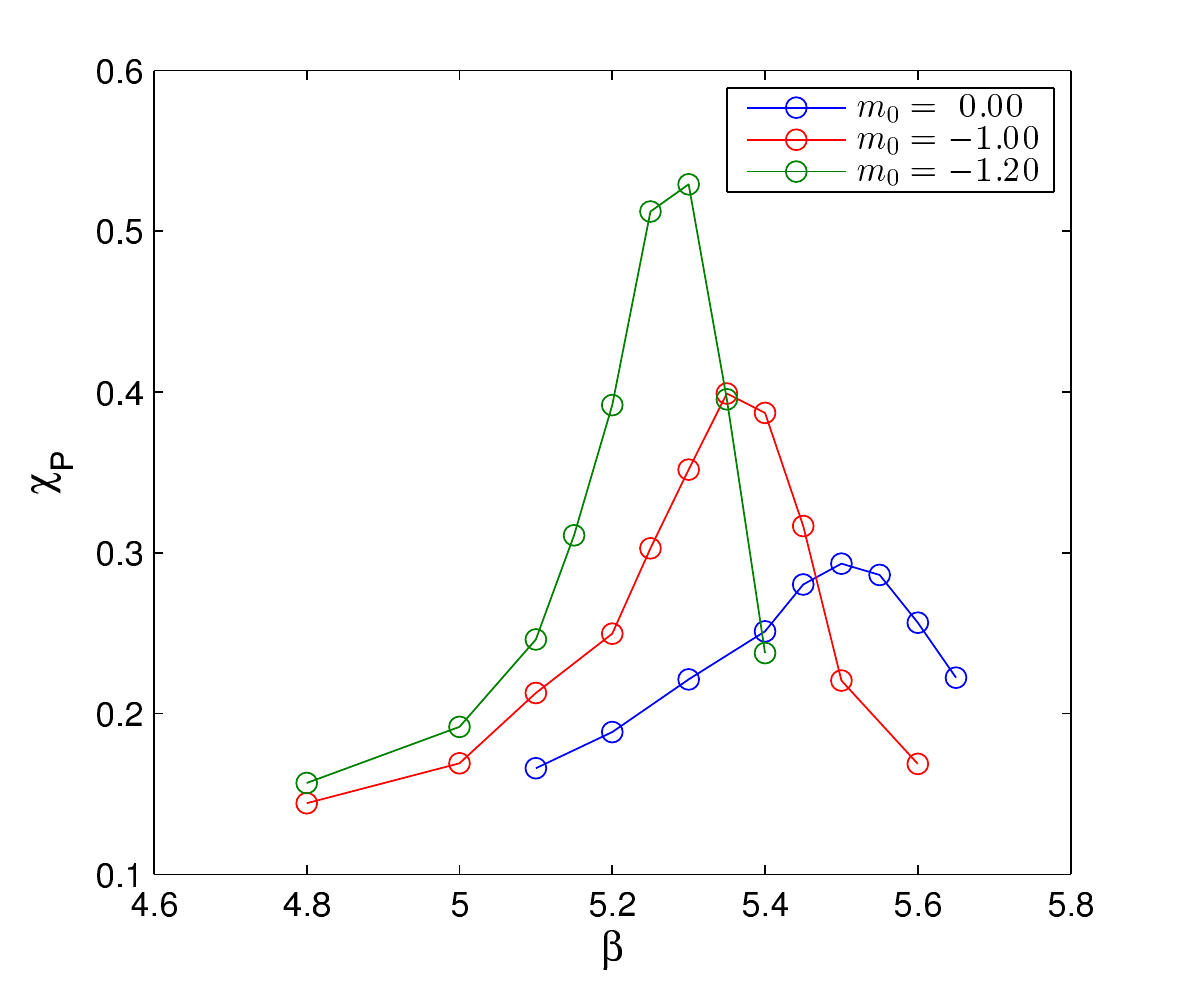}\\
    \includegraphics[height=5cm]{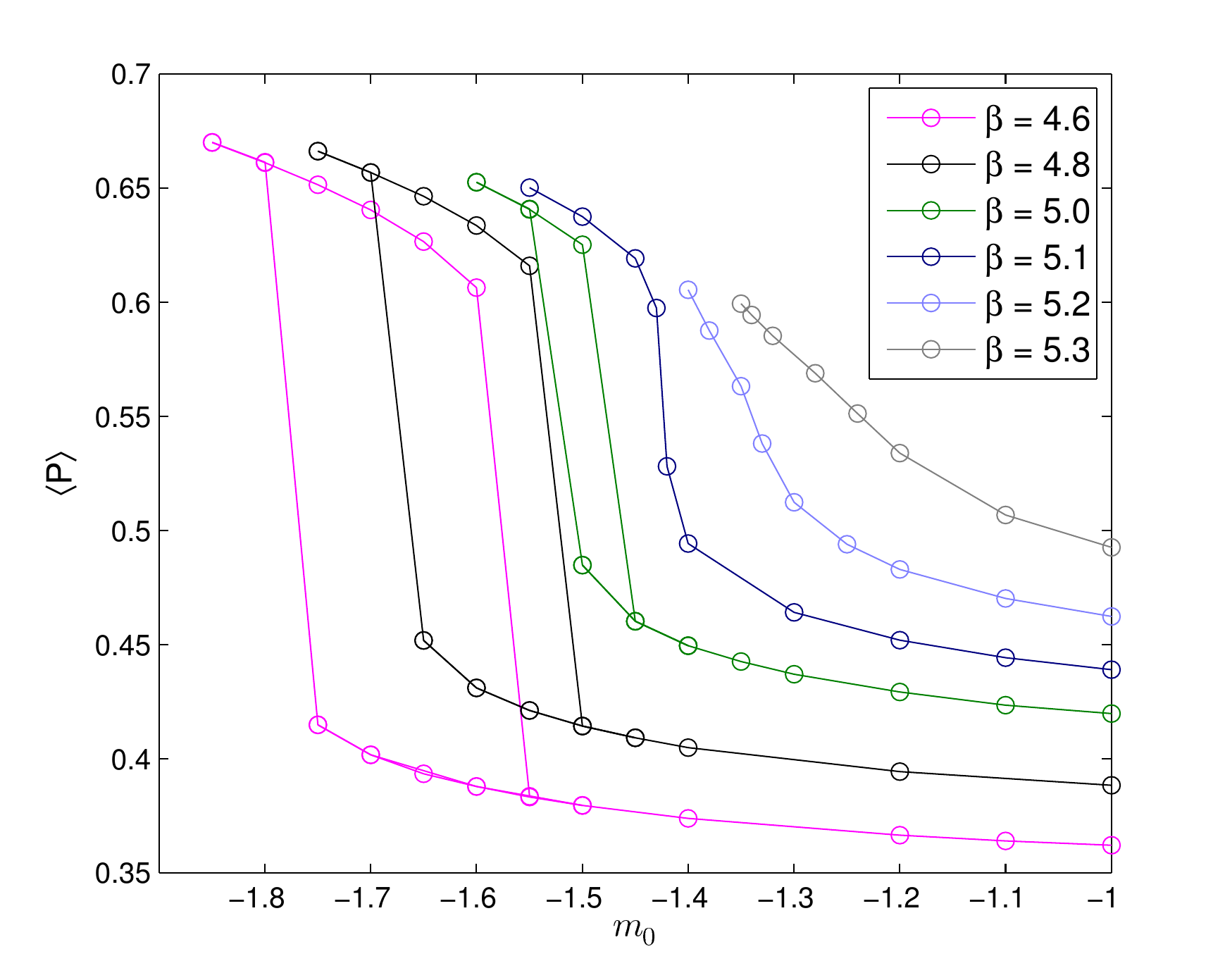}
    \includegraphics[height=5cm]{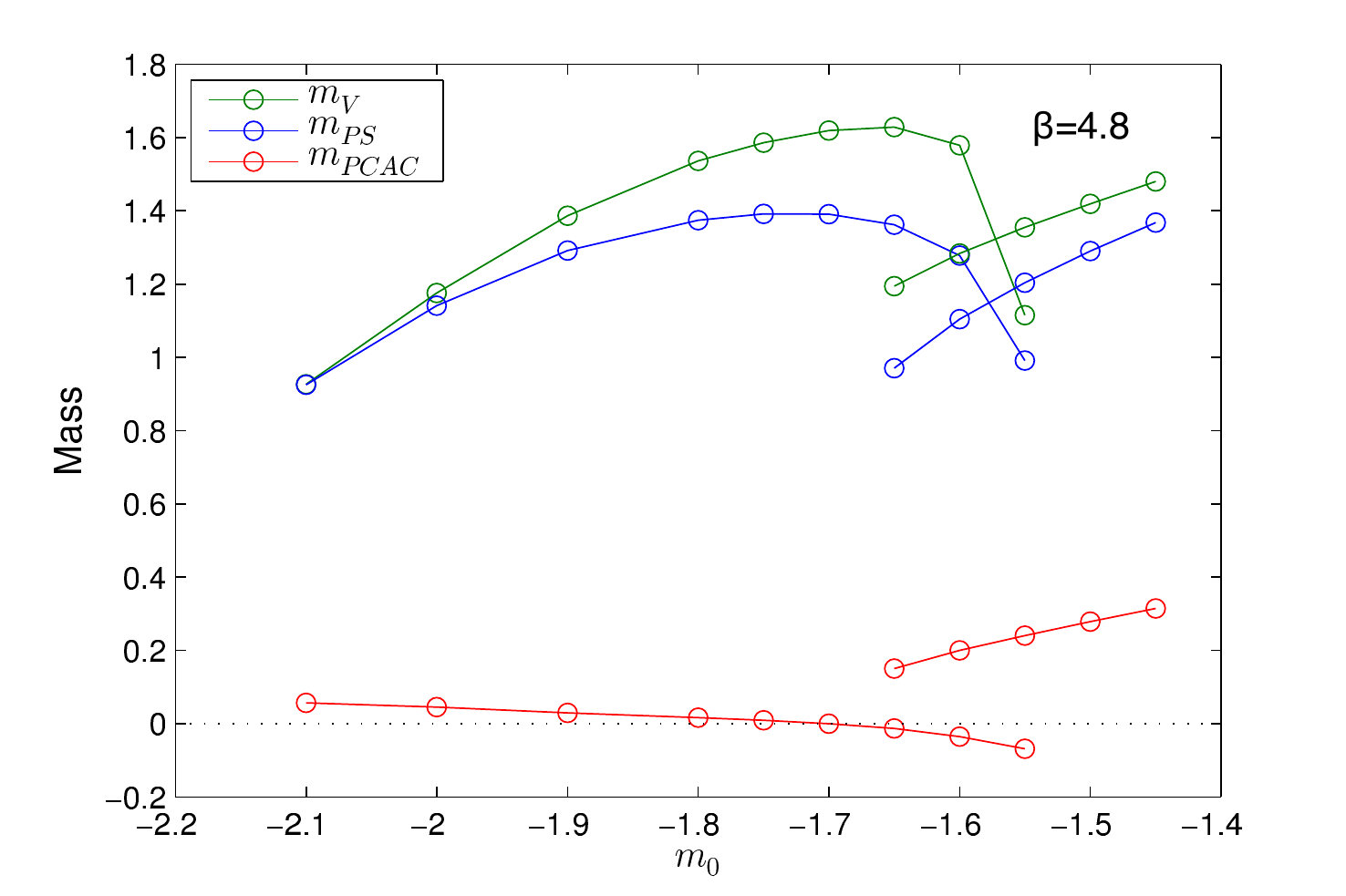}
  \end{center}
  \caption{(Top-left) Lattice phase diagram of the model investigated. (Top-right) The separation between a bulk phase (region I) and a weak coupling phase (region II) has been located by using the susceptibility of the plaquette. (Bottom-left) At strong coupling it is not possible to reach the zero quark mass limit as a first order transition is present, signalled by a strong hysteresis in the average plaquette when moving from region I to region IV. (Bottom-right) The same hysteresis can be observed in spectral observables.}
  \label{fig:pd}
\end{figure}
\section{Lattice Phase Diagram}\label{sec:pd}
Understanding the phase diagram of the latticed model is a crucial first step in the numerical analysis of a new model. The detailed phase diagram depends on the lattice action used, here the SU(3) plaquette action with non-improved Wilson fermions in the sextet representation.
We performed an extensive scan, comprising more than 200 simulations, in the parameter space of the bare coupling $\beta$ and the bare quark mass $m_0$ on $8^4$ or $16^3\times32$ lattices, depending on the observable, and used $24^3\times48$ lattices to check for finite volume effects.
We show in Fig.~\ref{fig:pd} the resulting phase diagram with four interesting regions. A strong coupling ``bulk'' phase is present separated from a weak coupling phase by a sharp crossover which we locate by looking at the susceptibility of the average plaquette (transition between region I and II in Fig.~\ref{fig:pd}).
When decreasing the bare quark mass at strong coupling, a (strong) first order transition can be observed by the associated hysteresis both in the average plaquette and spectral observables (bottom panels in Fig.~\ref{fig:pd}). This first order transition between region I and IV prevents the possibility to reach small quark masses at strong coupling.
At weaker coupling the the first order transition disappears and it is possible to reach into the chiral region with numerical simulations (transition between region II and III).

\begin{figure}
  \begin{center}
    \includegraphics[height=4.7cm]{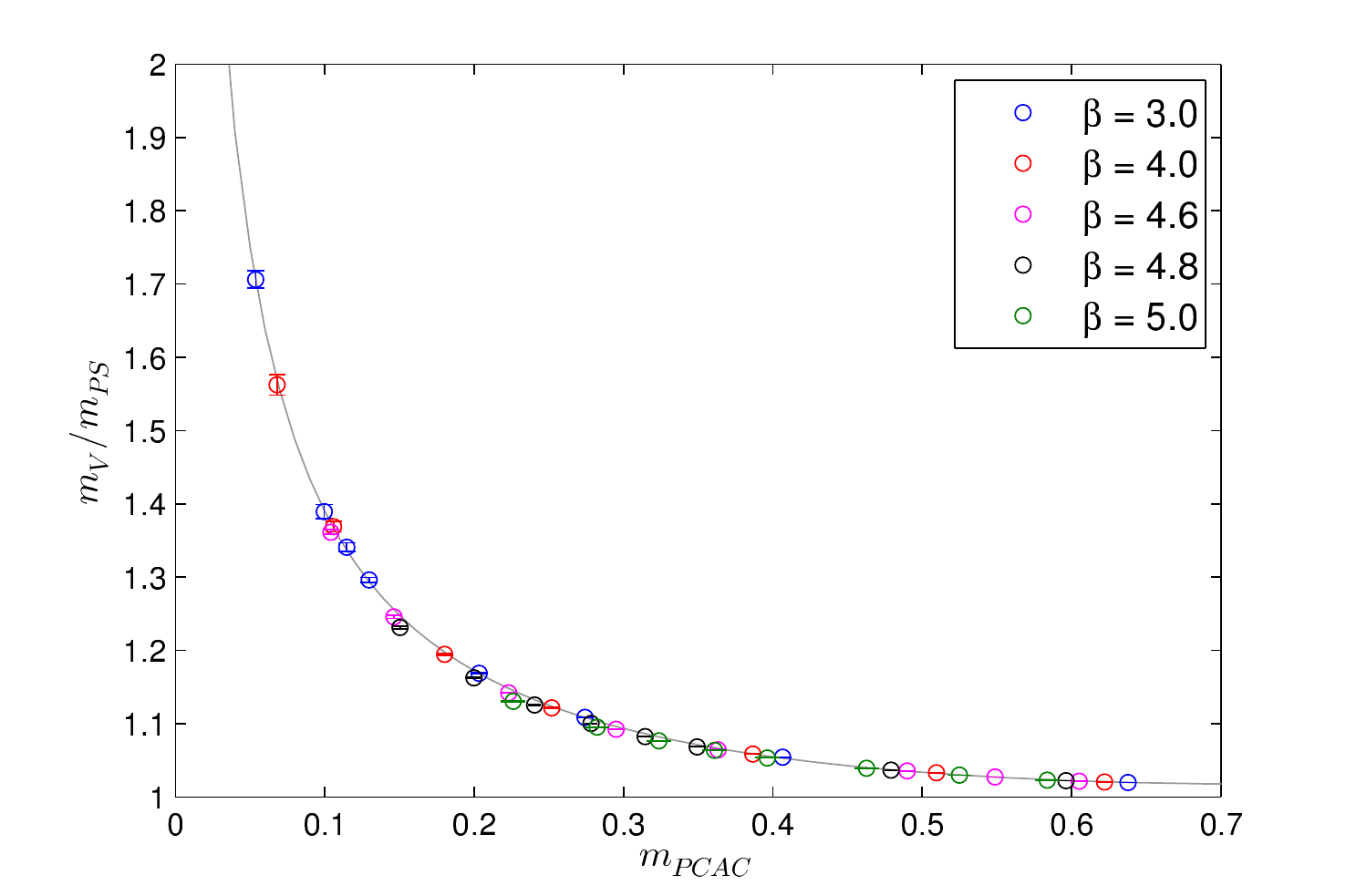}
    \includegraphics[height=4.7cm]{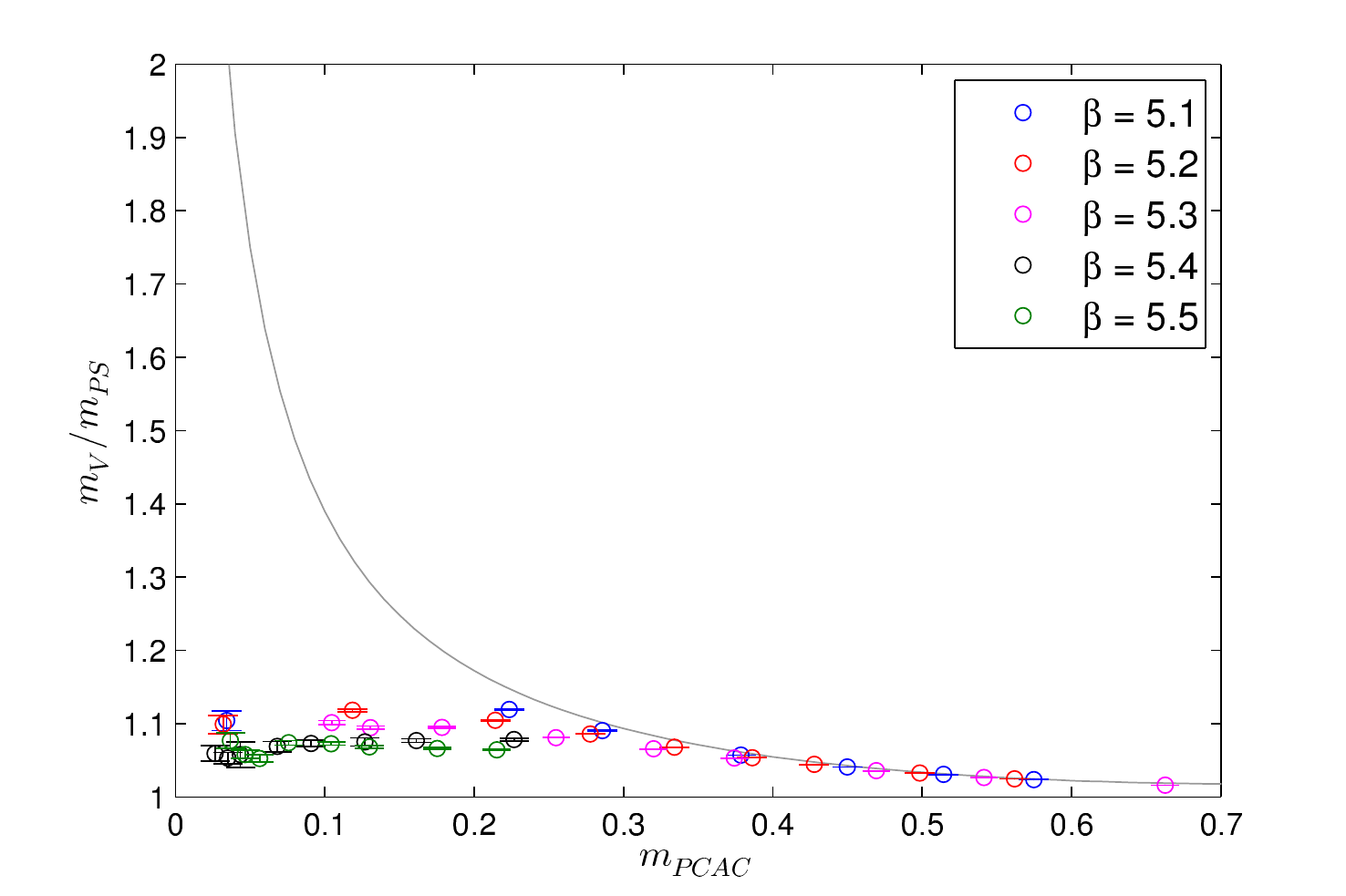}
  \end{center}
  \caption{Different behavior observed for the spectral quantity $m_V/m_{PS}$ as a function of the quark mass at strong coupling (left) and weak coupling (right).}
  \label{fig:spect1}
\end{figure}

\begin{figure}
  \begin{center}
    \includegraphics[height=4.7cm]{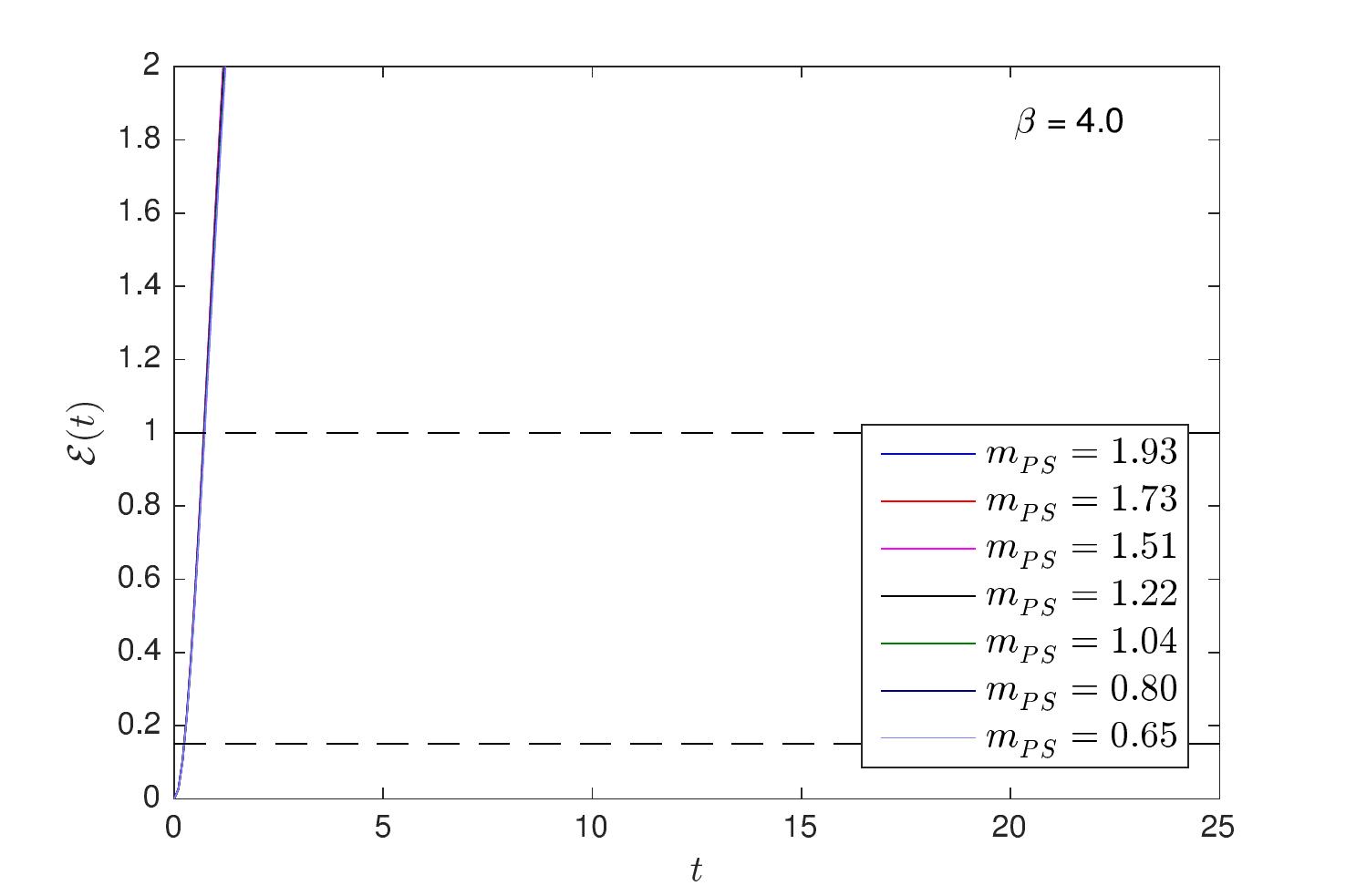}
    \includegraphics[height=4.7cm]{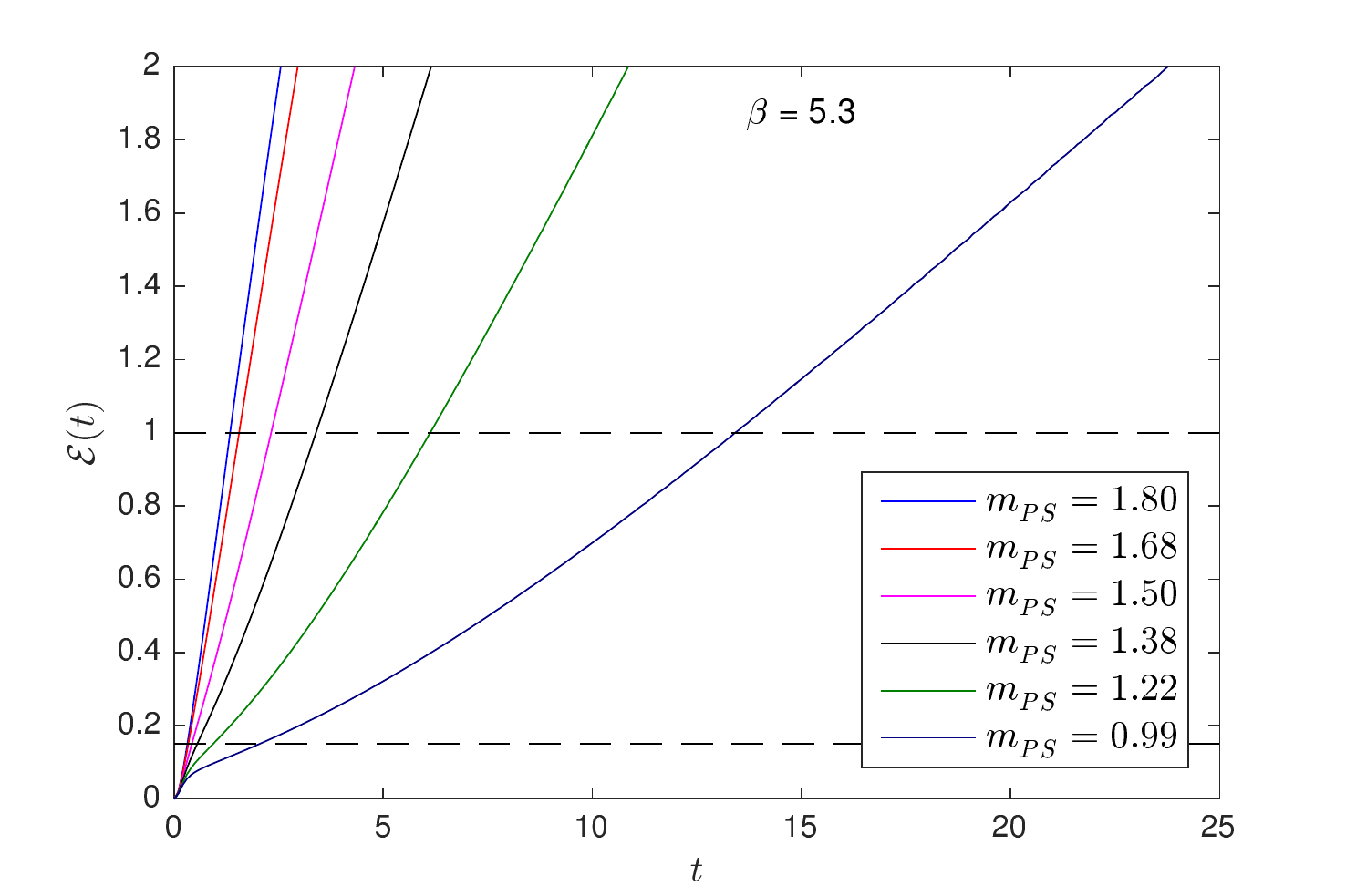}\\
    \includegraphics[height=4.7cm]{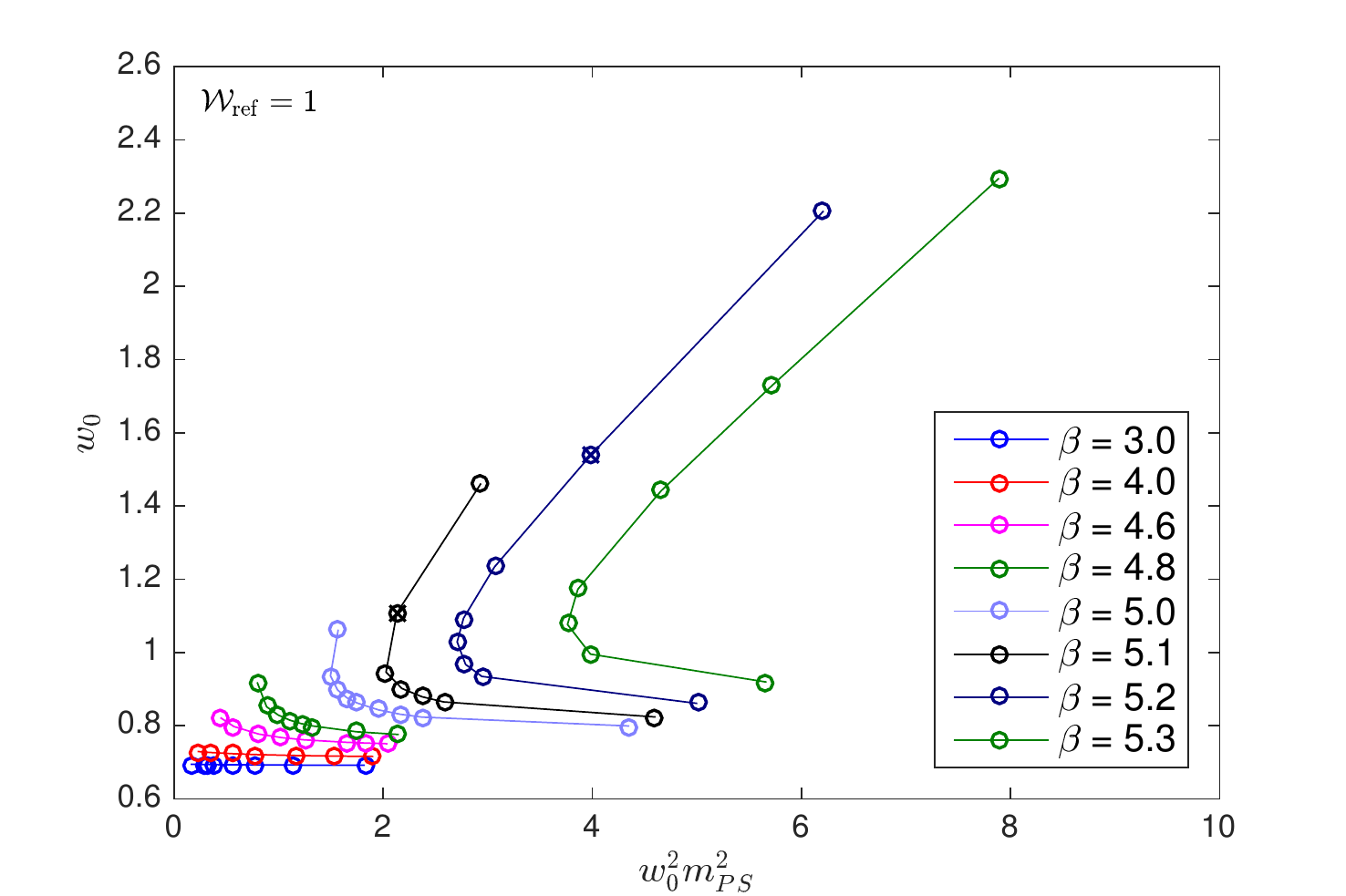}
    \includegraphics[height=4.7cm]{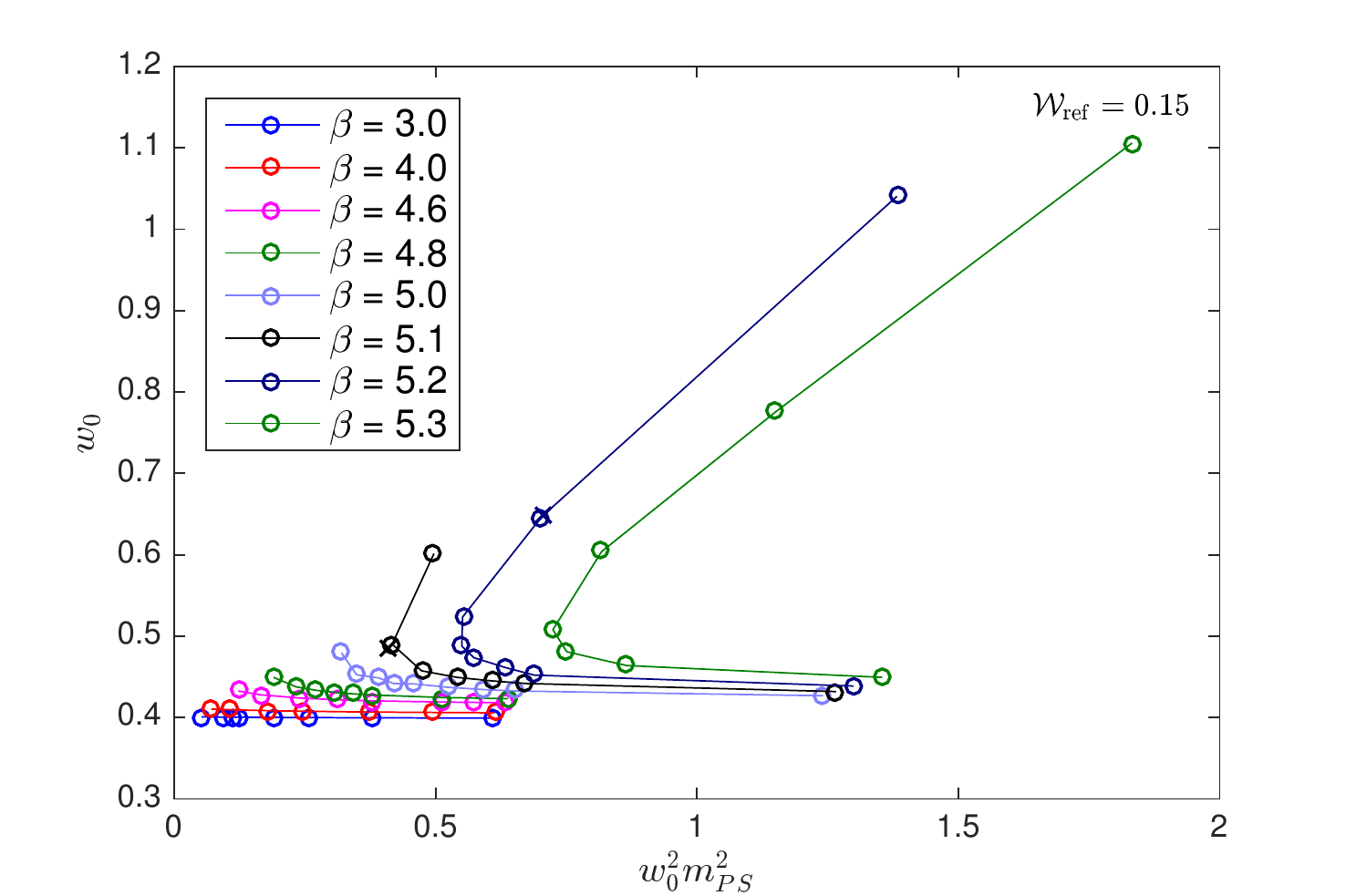}
  \end{center}
  \caption{(Top) Quark mass dependence of the gradient flow quantity $\mathcal{E}(t) = \langle t^2E(t)\rangle$ at strong coupling (left) and weak coupling (right). (Bottom) We used two different reference values to define a scale $w_0$ (equivalently $\sqrt{t_0}$). For both values, $w_0$ grows faster than $1/m_{PS}$ in the chiral limit in the weak coupling region. Finite volume effects are negligible.}
  \label{fig:gf}
\end{figure}

We studied the behavior of two interesting quantities near the light quark mass region both at strong and weak coupling (region I and II respectively).
We show in Fig.~\ref{fig:spect1} the quantity $m_V/m_{PS}$ as a function of the PCAC mass both in the strong (left panel) and the weak (right panel) coupling regions.
At strong coupling the ratio clearly increases towards the chiral limit and we observe a clear splitting between the vector and pseudoscalar meson masses, which is consistent with the expectation of chiral symmetry breaking at strong coupling.
However when moving to weak couplings in Region II (right panel) we observe that the ratio becomes almost constant towards the chiral limit and the two states remain almost degenerate over the entire range of quark masses investigated here. 
The second quantity investigated is the scale setting observable $w_0$ (equivalently $\sqrt{t_0}$) defined via the gradient flow, shown in Fig.~\ref{fig:gf}. At strong coupling this quantity only shows a modest dependence on the pseudoscalar meson mass $m_{PS}$, similarly to the case of QCD where a reliable extrapolation to the chiral limit is possible. On the other hand at weak coupling, we observe a very strong dependence on $m_{PS}$. In fact at weak coupling $w_0$ grows faster than $1/m_{PS}$ for light quark masses, which is not the expected behavior in a chirally broken model.

We stress that finite volume effects have been extensively checked for our data  and the results presented here should be understood as the ``infinite volume'' behavior of the lattice model. For example in Fig.~\ref{fig:gf}, in the bottom row, we include the results from a larger $24^3\times 48$ volume (cross symbols), which lie on top of the smaller volume results within errors. Results for the spectrum on even larger volumes are presented below.


\section{Spectrum in the weak coupling region}\label{sec:weak}
We now focus on the weak coupling region and study the chiral behavior of the model at one fixed value of the lattice spacing, corresponding to $\beta=5.4$. 
We measure the quantities $m_{PCAC}$, $f_{PS}$, $m_{PS}$, $f_{V}$, $m_V$ and the mass of a baryonic state $m_N$. We use volumes up to $32^3\times 48$ to control finite volumes effects, which were found to be negligible compared to statistical errors for all data presented here. The detailed setup for the numerical computation and the finite volume analysis can be found in \cite{Hansen:2017ejh}.

\begin{figure}
  \begin{center}
    \includegraphics[height=5.5cm]{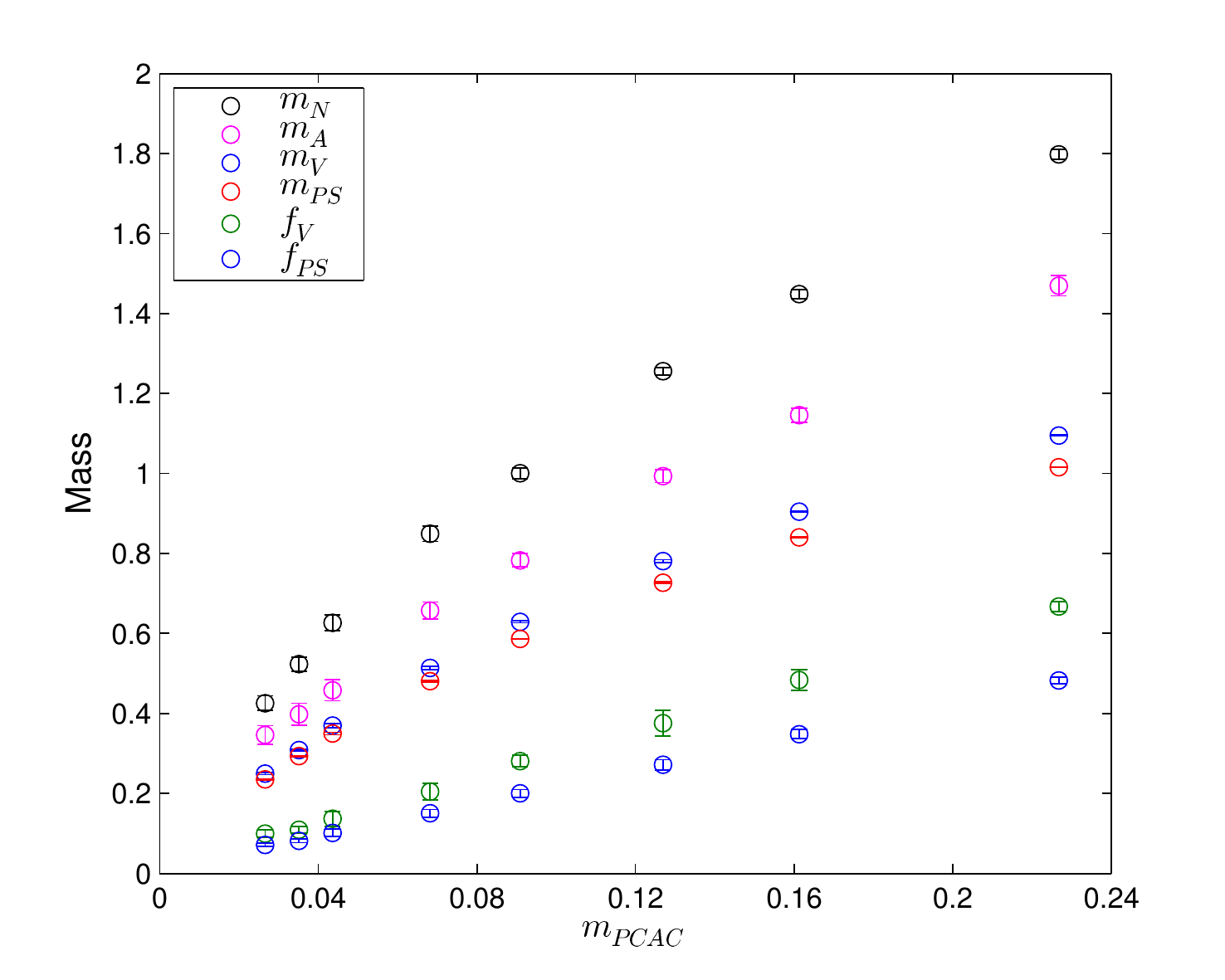}
    \includegraphics[height=5.5cm]{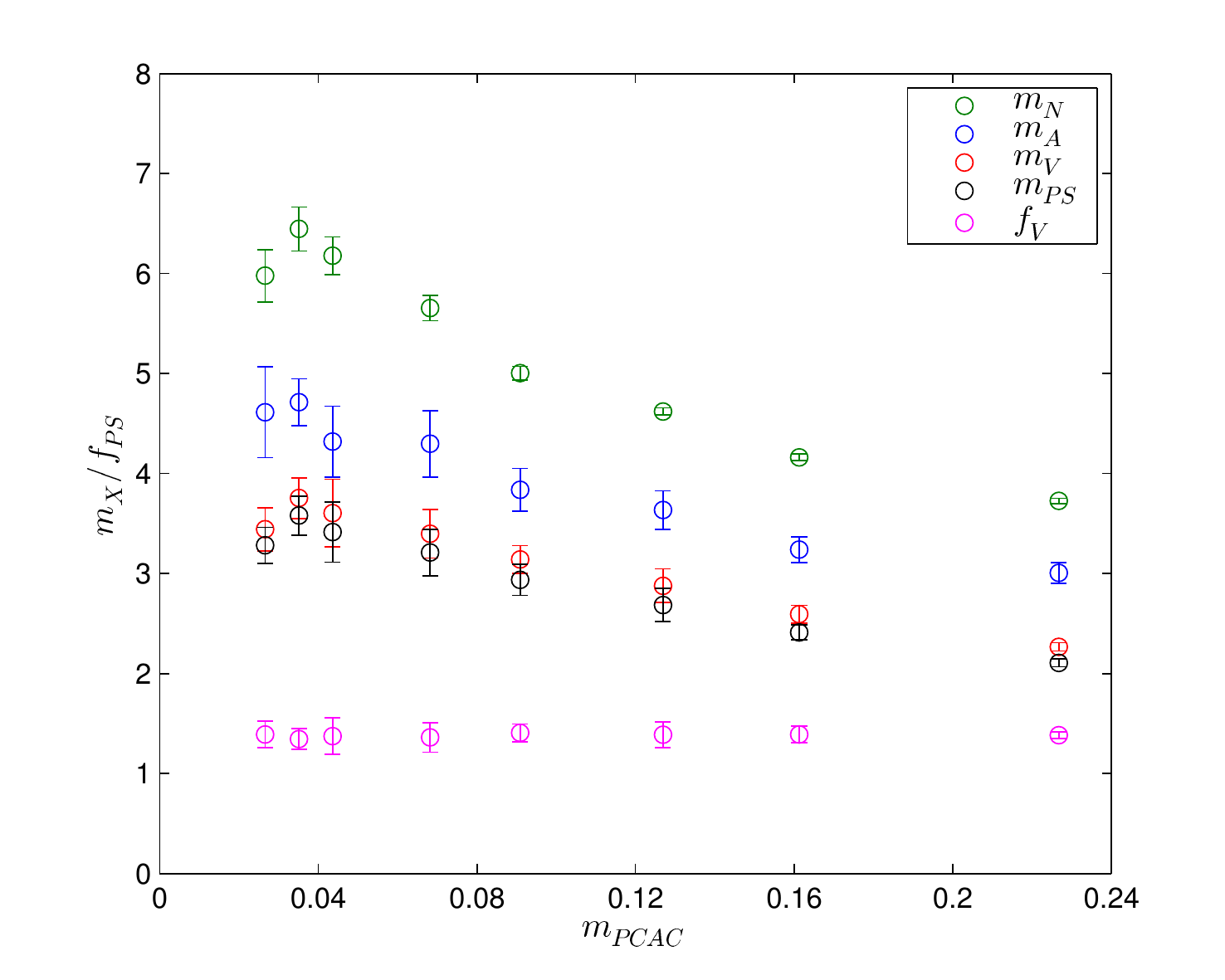}
  \end{center}
  \caption{(Left) Measured spectrum of the model as a function of the PCAC quark mass. (Right) Same spectral quantities normalized by $f_{PS}$.}
  \label{fig:spect2}
\end{figure}

A summary picture of all measured spectral quantities is shown in the left panel of Fig.~\ref{fig:spect2}, whereas in the right panel we normalize the same quantities by $f_{PS}$, in both cases as a function of the PCAC quark mass.
We observe that all the spectral quantities considered here have a strong dependence on the quark mass and the ratios to $f_{PS}$ seem to converge to a finite non-zero value in the chiral limit. 
A similar behavior has been observed in all previous lattice studies of the spectrum of conformal and near conformal models, including the sextet model considered here~\cite{Fodor:2016pls}.   
If such a behavior would persist to arbitrarily small quark masses, the theory would be IR conformal.

\begin{figure}
  \begin{center}
   \vspace{-3mm}
   \includegraphics[width=0.45\textwidth]{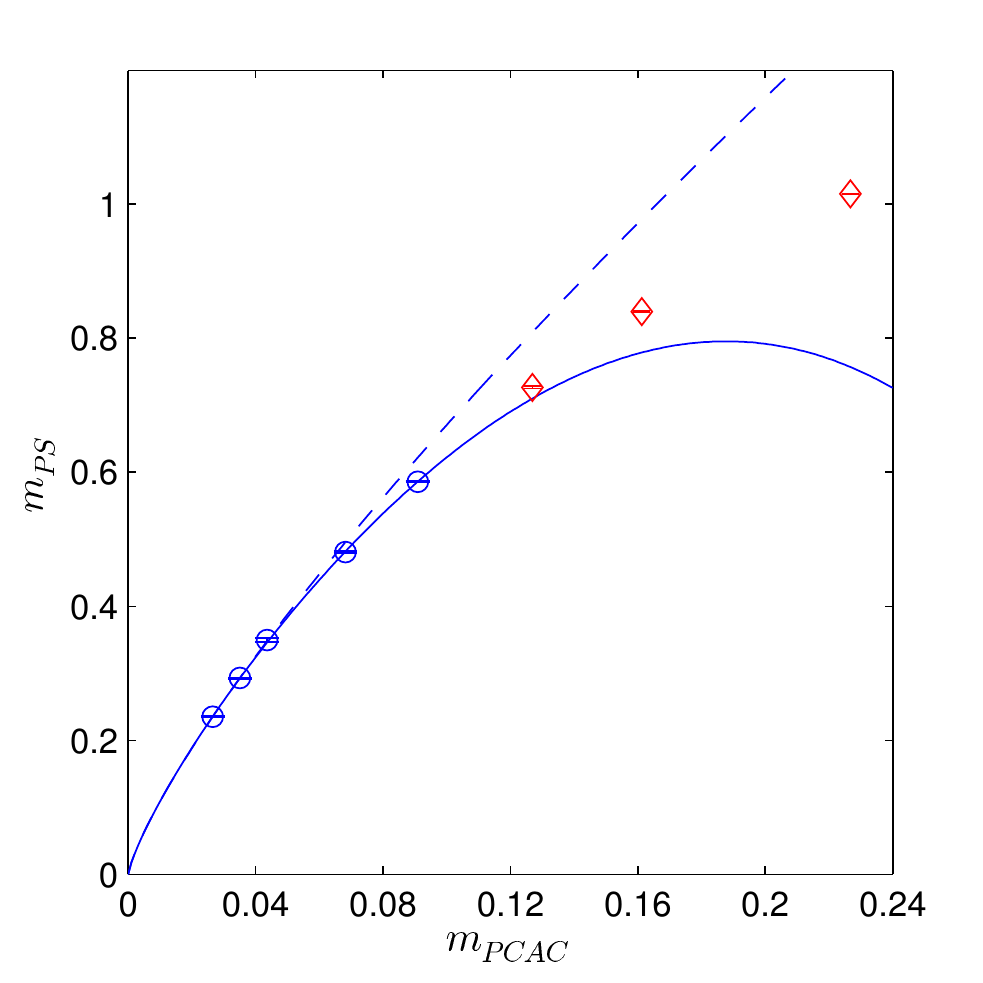}
   \includegraphics[width=0.45\textwidth]{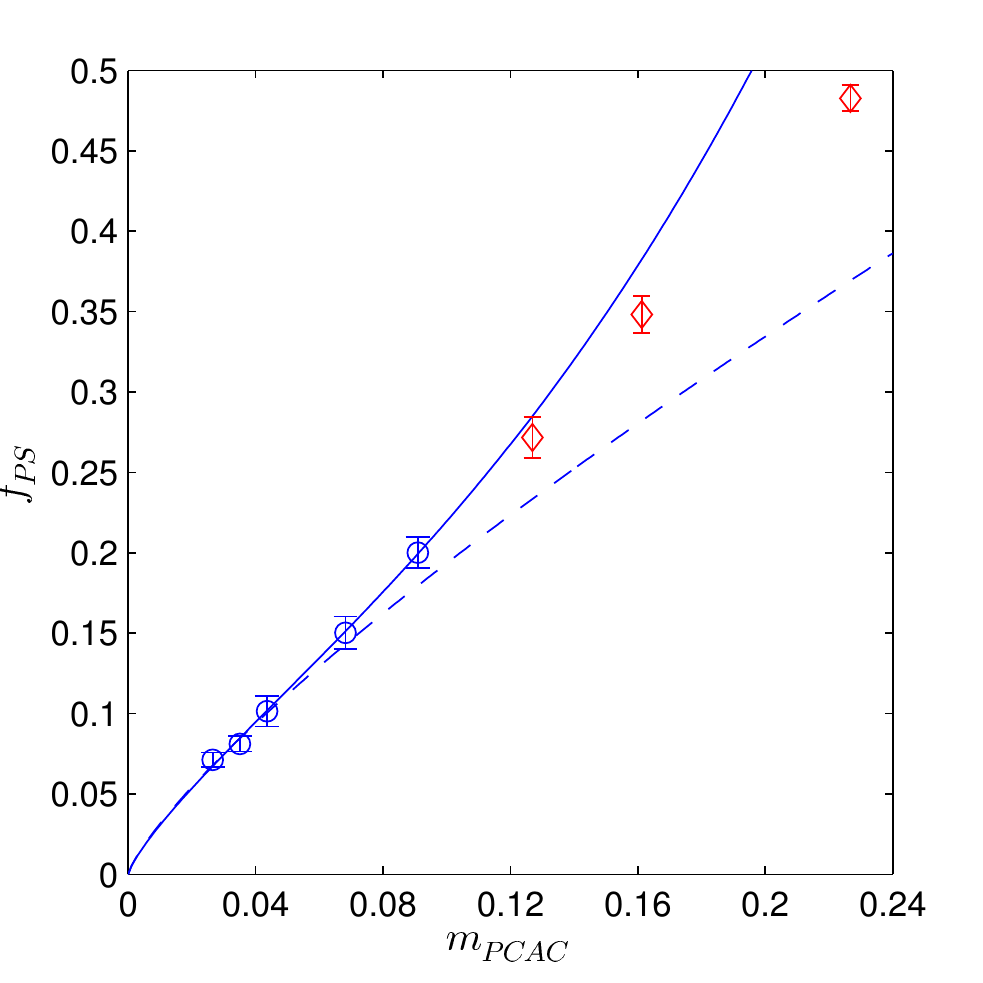} \\
   \vspace{-3mm}
   \includegraphics[width=0.45\textwidth]{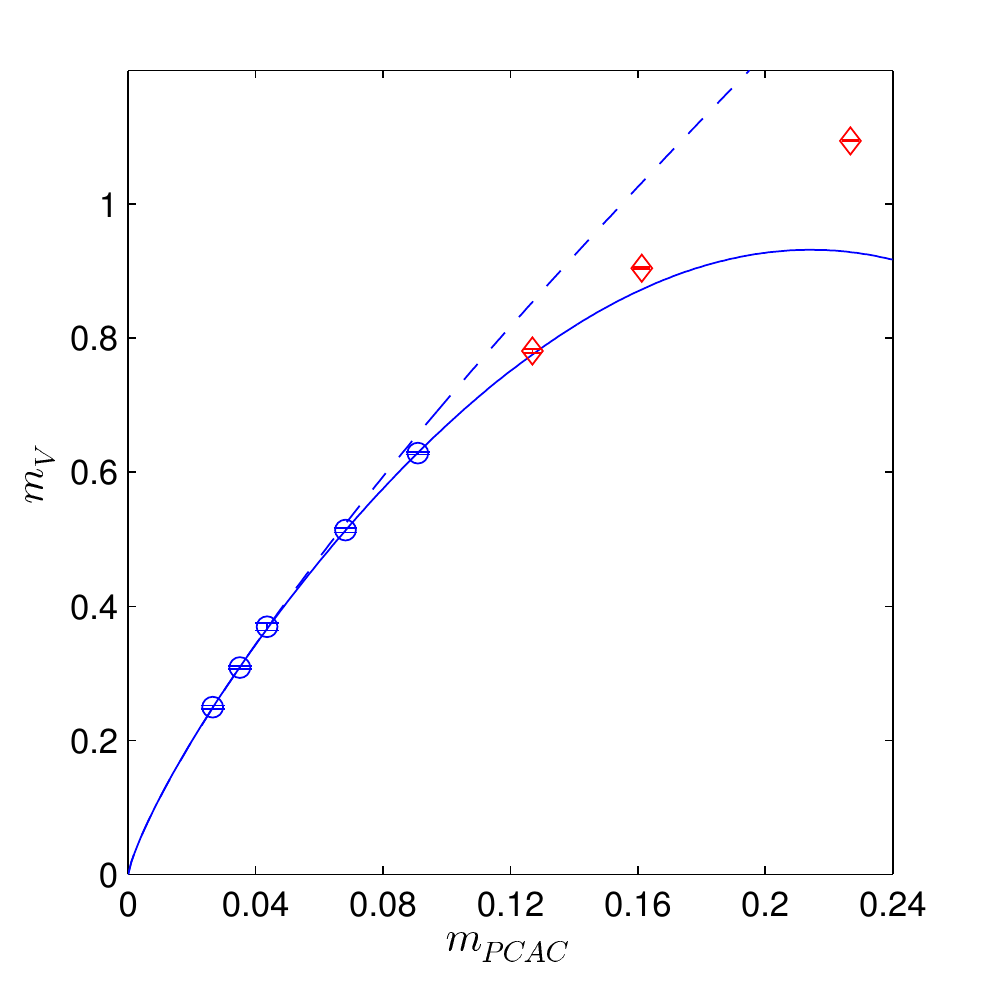}
   \includegraphics[width=0.45\textwidth]{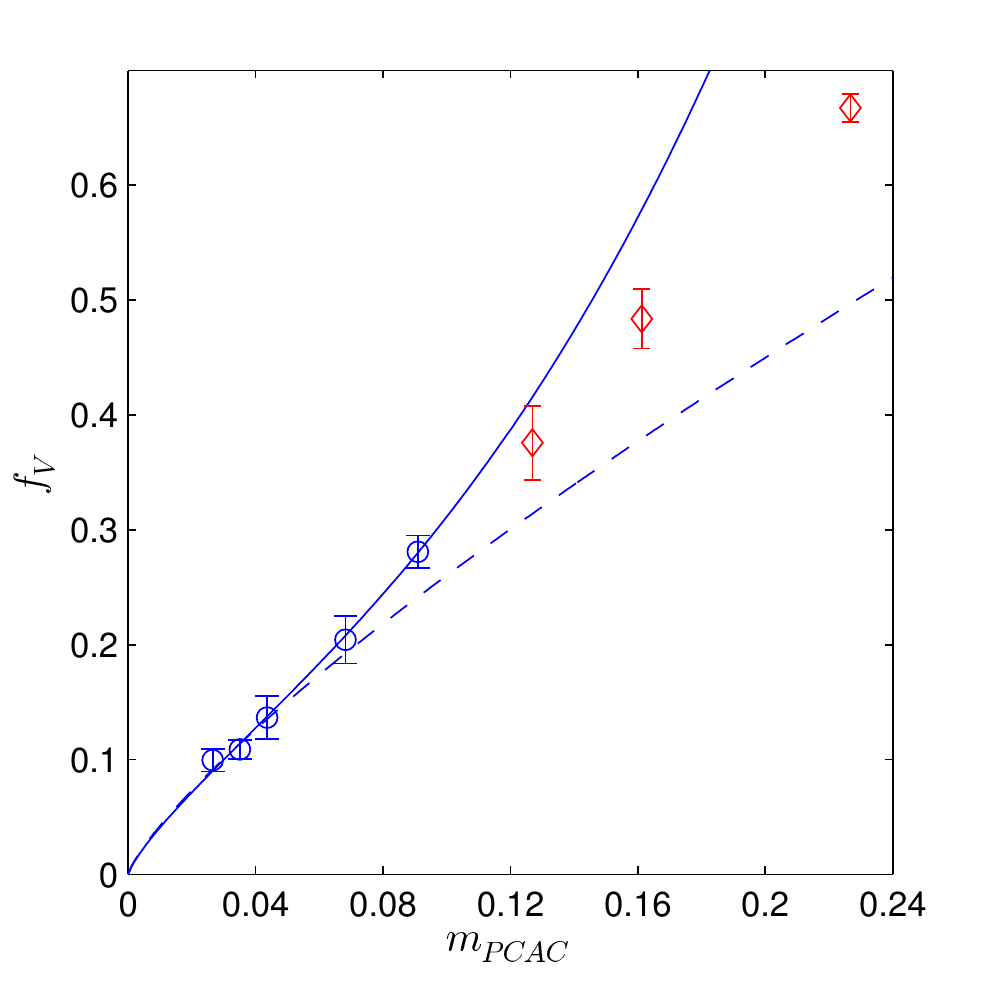} \\
   \vspace{-3mm}
   \includegraphics[width=0.45\textwidth]{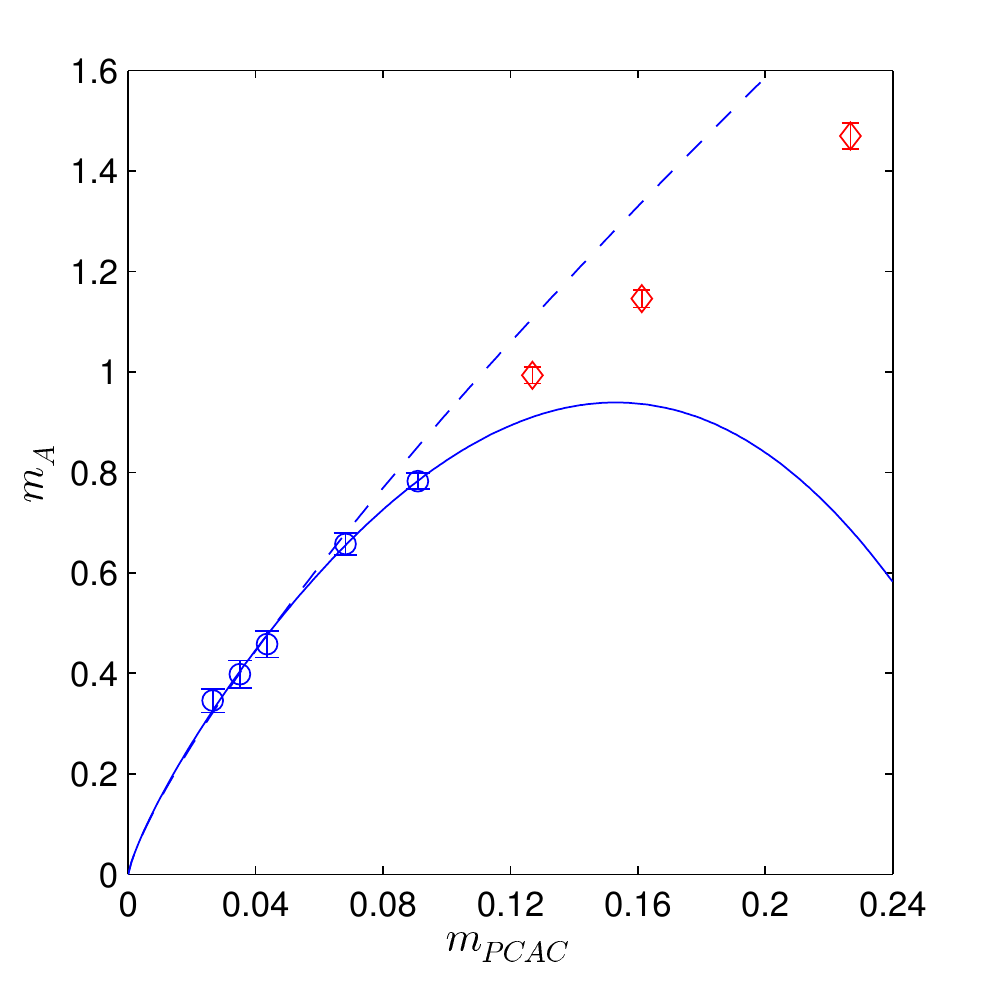}
   \includegraphics[width=0.45\textwidth]{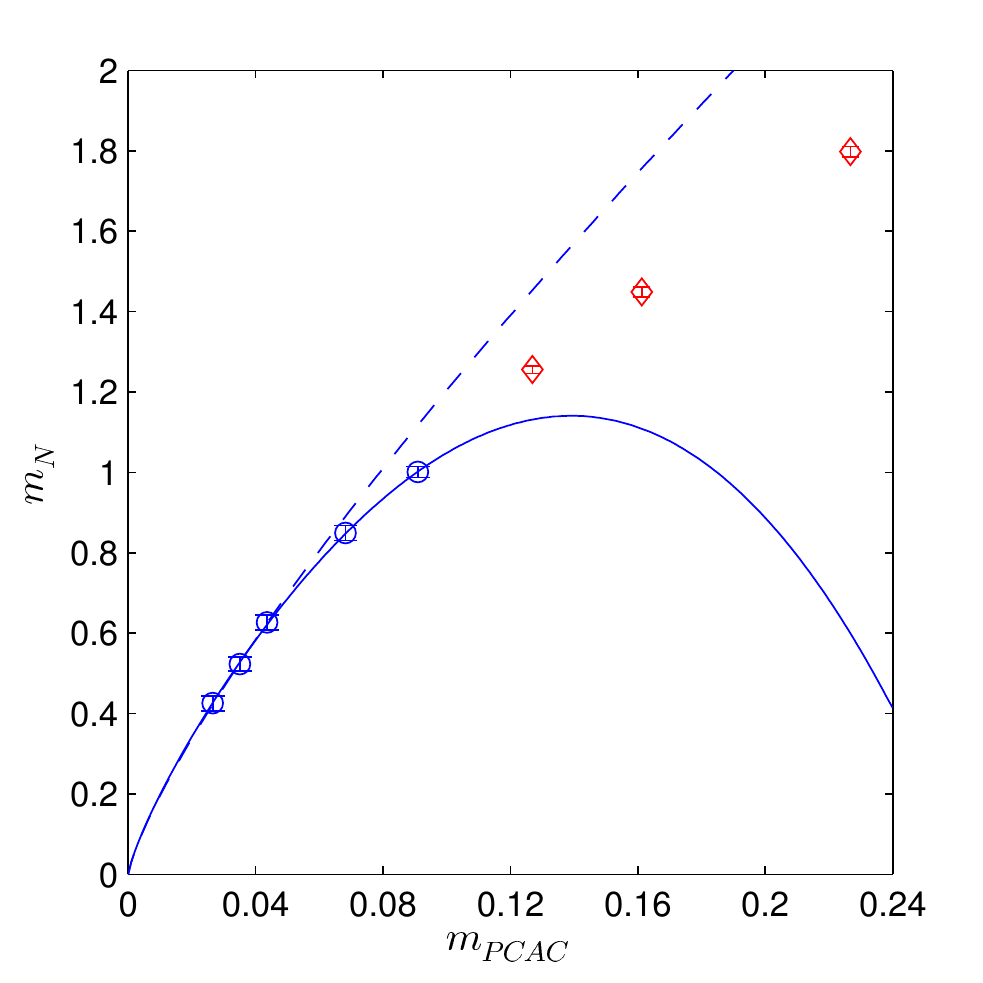} \\
   \vspace{-3mm}
  \end{center}
  \caption{Combined conformal fits for $\beta=5.4$ with and without subleading corrections. 
  The dashed line is the leading order conformal fit to the three lightest points in all channels. The solid line is an independent fit with subleading scaling corrections, to the five lightest points in all channels.}
  \label{fig:conformal}
  \end{figure}

To obtain a quantitative statement about the chiral behavior of the model, we have fitted the measured spectrum to the analytical expectations available for both the case of IR conformality and chiral symmetry breaking.
To test for signals of IR conformality, we performed a single, combined fit of all available spectral quantities which includes the leading order scaling behavior and the first sub-leading correction to scaling. The result for the combined fit is shown in Fig.~\ref{fig:conformal}.
The fit (solid line) describes the data well over a range of quark masses up to $m_{PCAC}\sim 0.10$ with a $\chi^2/\mathrm{dof}=7.04/16=0.44$.
The best fit value for the anomalous dimension of the mass is $\gamma=0.25(3)$ and the subleading exponent is $\omega=2.71(76)$.
As a consistency check, we also performed the same combined fit, but excluding subleading corrections to scaling in a narrower range of masses.
The fit to the leading order behavior (dashed line) works well for masses $m_{PCAC}<0.05$ with $\chi^2/\mathrm{dof}=4.62/11=0.42$.
The best fit value for the mass anomalous dimension extracted from this fit is $\gamma=0.27(3)$, which is in very good agreement with the value obtained considering corrections to scaling.
We conclude that the IR conformality hypothesis is in good agreement with our present dataset.

To test for chiral symmetry breaking turns out to be more problematic, as an analytical understanding of some quantities, like $m_{PS}$ or $f_{PS}$ as a function of the quark mass, is needed over the accessible range of quark masses.
In the case of QCD one would rely on chiral perturbation theory to make contact with the chiral limit. 
However in the present case the applicability of chiral perturbation theory is very doubtful. 
By inspection, one immediately realizes that the vector meson state is almost degenerate with the would-be Goldstone boson, and that $f_{PS}$ is rapidly changing with the PCAC mass. 
Moreover there are indications from simulations performed with staggered fermions that a light scalar state could be present in the model~\cite{Fodor:2016pls}, in fact even lighter than $m_{PS}$ in the mass range within the reach of lattice simulations.

\begin{figure}
    \begin{center}
      \includegraphics[height=5.5cm]{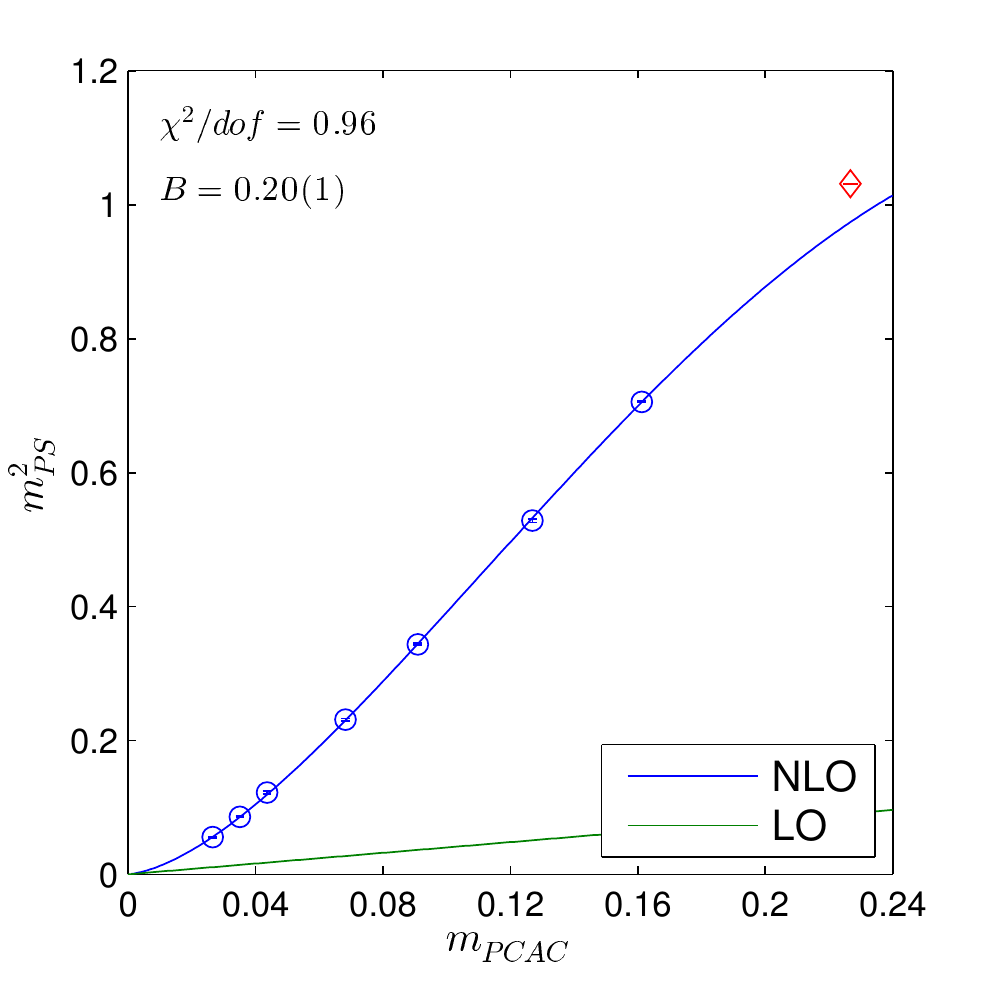}
      \includegraphics[height=5.5cm]{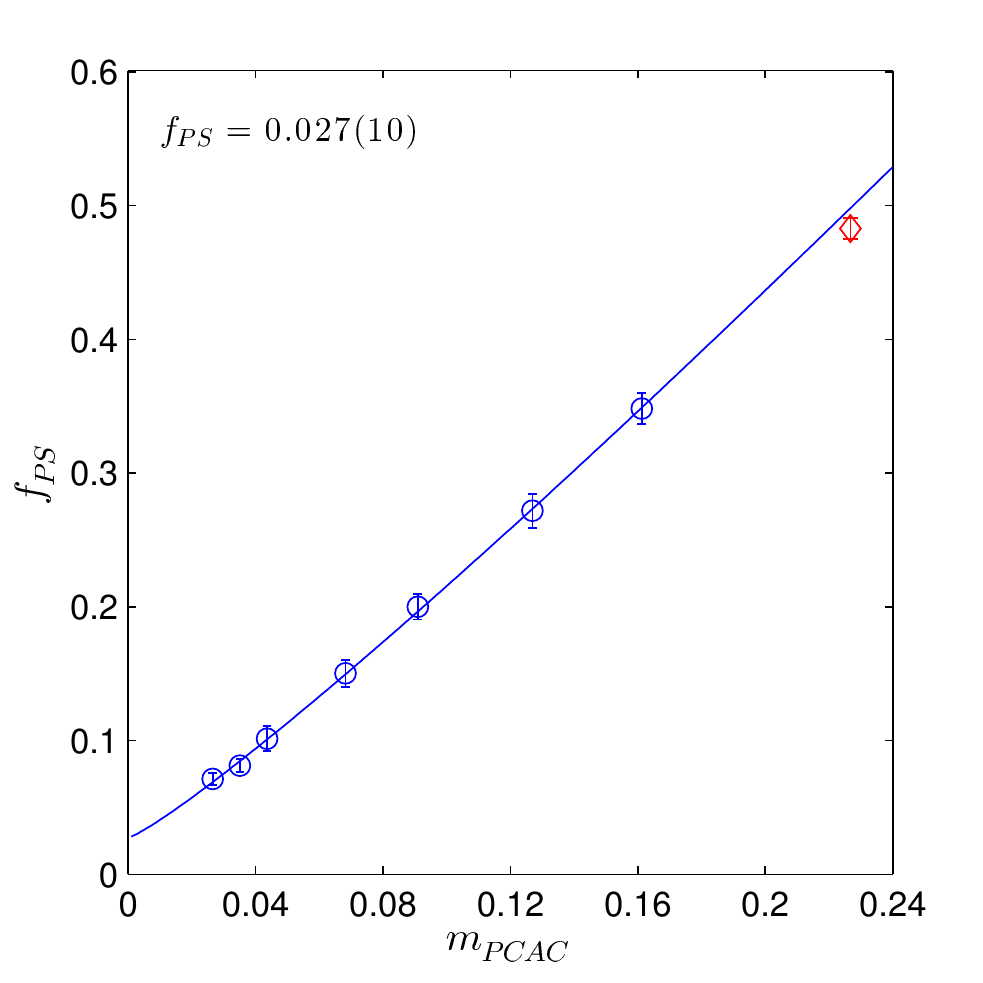}
    \end{center}
    \caption{Best fit to a functional form inspired by chiral perturbation theory, as described in the text.}
    \label{fig:cpt}
  \end{figure}

While several proposals exists for an alternative effective description that could maybe apply to the present case, there is no compelling choice yet among them.
Here we resolve, as a practical solution, to use several functional forms for $m_{PS}$ and $f_{PS}$ inspired by chiral perturbation theory, both in the continuum or for Wilson fermions.
Specifically we use 
\begin{align}
  \begin{split}
   m_{PS}^2 &= M^2 \left[ 1 + \frac{M^2}{F^2}(a_ML+b_M) + \frac{M^4}{F^4}(c_ML^2+d_ML+e_M) \right]\,,\\
   f_{PS} &= F \left[ 1 + \frac{M^2}{F^2}(a_FL+b_F) + \frac{M^4}{F^4}(c_FL^2+d_FL+e_F)\right] \,,
  \end{split}\label{eq:cpt}
  \end{align}
where $M^2=2Bm$ is the leading order pion mass from the GMOR relation, $F$ is the leading order pion decay constant, and $L$ is short-hand notation for the chiral logs:
\begin{equation}
 L = \frac{1}{16\pi^2}\log\left(\frac{M^2}{\mu^2}\right)\,.
\end{equation}
The coefficients $\{a_M,a_F,c_M,c_F\}$ are known for the continuum chiral perturbation theory under consideration, but can also be left as free parameters.
We performed several fits to our numerical data for $m_{PS}$ and $f_{PS}$, both at NLO and NNLO, with or without fixing the contribution of the chiral logs.
A typical example is in Fig.~\ref{fig:cpt}, where we show the best fit to our data at NLO with free coefficients for the chiral log terms. 
Although the fit describes well the available data over a large range of quark masses, we find two main features common to all acceptable fit functions: 1) higher order corrections are very dominant over the entire range of quark masses explored; 2) the best fit functions tend to be simple polynomials in the quark mass -- in fact both $f_{PS}$ and $m_{PS}$ are largely simple linear fuctions of $m_{PCAC}$.

In view of these results, althought it is possible to use a functional form inspired by chiral perturbation theory to describe our numerical data, we conclude that little physical meaning can be extracted from these fits.
If the model is chirally broken, substantially lighter quark masses are needed to make contact with ChPT.

\section{Conclusions}\label{sec:conclusions}
In this work we presented our results for the infrared properties of the SU(3) ``sextet'' model by focusing on the physical spectrum.  
Most results available for the spectrum of this model were obtained with staggered fermions. Here we chose the Wilson discretization for a comparison.
We mapped out the phase diagram of the lattice model and located several distinct regions in the parameter space, see Fig.~\ref{fig:pd}, which show very different qualitative behaviors. In particular, while we observe clear qualitative signals of chiral symmetry breaking in the strong coupling region, those are absent in the weak coupling region.
We then investigated the weak coupling region in more detail, aiming to reach as light quark masses as possible, while keeping finite volume corrections under control. 
Our results for the spectrum of the model in the weak coupling phase are compatible with the IR conformal scenario. If chiral symmetry breaking occurs and the model is very slowly ``walking'', then much lighter quark masses than currently accessible will be needed to probe into the chirally broken dynamics.

This work was supported by the Danish National Research Foundation DNRF:90 grant and by a Lundbeck Foundation Fellowship grant. The computational resources were provided by the DeIC national HPC centre at SDU.

\bibliography{lattice2017}

\end{document}